\documentclass{article}
\usepackage{graphics,epsfig}
\usepackage{times,cite}
\usepackage[english]{babel}
\begin{document}

\title{Statistical analysis of neutrino events from SN1987A
neutrino burst: estimation of the electron antineutrino mass}

\author{B. I. Goryachev,\\
Skobeltsyn Institute of Nuclear Physics,\\
Lomonosov Moscow State University,\\
119991, Moscow, Russia,\\
e-mail: bigor@srd.sinp.msu.ru
}

\maketitle

\begin{abstract}
The method of the statistical sample moments was
used for the analysis of neutrino events from SN1987A burst in
Cherenkov detectors. In particular the coefficients of correlation
$Q(E,t)$ between the energies $E$ of electron antineutrinos $\bar \nu_e$
emitted by star and the ejection instants $t$ of $\bar \nu_e$ for neutrino
events recorded by Cherenkov water detectors of KII and IMB collaborations
were calculated. $Q(E,t)$ values depend on the assumed mass of $\bar \nu_e$.
Modern model of the gravitational stellar core collapse with an accretion 
phase predicts the low level of  such correlation $\langle Q(E,t)\rangle$
averaged with respect to neutrino burst. On condition that empirical $Q(E,t)$
values equal the theoretical model quantities $\langle Q(E,t)\rangle$ one can
obtain $22\pm10\,eV/c^{2}$ as an estimate of the nonzero $\bar \nu_e$
mass.  The error of this estimate implies  that $\bar \nu_e$ mass less than 
$2 eV/c^2$ is unlikely. The laboratory data of the tritium $\beta$-decay
agree adequately with the presented astrophysical estimate provided that
the anomalous structure near the end point of $\beta$-spectrum is taken
into account.
\end{abstract}

\section{Introduction}

The observation of SN1987A supernova
neutrino burst gave valuable information related to the star collapse
astrophysics and neutrino physics. Mainly this information was obtained
by Cherenkov water detectors of Kamiokande II (KII)~\cite{1} and
Irvine-Michigan-Brookhaven (IMB)~\cite{2} collaborations. Two decades
before the burst, Zatsepin suggested the possibility of estimation of
the neutrino mass taking into account that massive neutrino with rest
energy $E_0$ and total energy $E$ passes distance $d$  from the source to
detector
in a time~\cite{3}

\begin{equation}
\label{eq1}
t_{sd} = \frac{d}{c}(1+\frac{1}{2}\frac{E_0^2}{E^2})
\end{equation}

Here $c$  is the velocity of light. Large
number of works was dedicated to the estimation of $E_0$ beginning from
1987, in which as a rule, the parameters of individual neutrino events,
namely $E$  and the time of neutrino registration $t_d$  were compared. Such
analysis has large uncertainty due to the finite duration of neutrino
burst. An opinion was formed that the results of observations agree
with zero neutrino mass or allow one to estimate only its upper
limit.

\section{Statistical analysis}
In this paper the results of estimation
the electron antineutrino $\bar \nu_e$ mass by comparison of the lower
empirical
sample central moments of two dimensional density function $F(E,t)$ of the
source and of similar moments of this function predicted by a modern
theoretical model~\cite{4} are presented
\footnote{ Below $F(x,y)$ and $Q(x,y)$ stand for respectively
two dimensional density function and the
coefficient of correlation of variables $x$ and $y$. Furthermore, $D(z)$ and
$S(z)$ are the variance and the standard deviation of quantity $z$
respectively.}.
Empirical moments of $F(E,t)$ depend on the assumed value of $E_0$
because $t_d = t+t_{sd}$, where $t$ is the instant of the
$\bar \nu_e$ emission from the source. As
usual, the energy of neutrino during its motion from the source to
detector is assumed to be constant.

Density function at distance $d$ from the
source is

\begin{equation}
\label{eq2}
F(E,t_d)=\int_0^\infty\delta(t_d-t-t_{sd})F(E,t)dt
\end{equation}

Here zero time $(t=0)$ corresponds to the
beginning of emission of $\bar \nu_e$.
If one considers a specific detector, it is
convenient to normalize $F(E,t)$ in formula (\ref{eq2})
to unity in domain $0<t<\infty$ and $E_c \le E < \infty$.
Here $E_c$  designates energy threshold of  $\bar \nu_e$
detection in considered detector. All sample empirical moments of
$F(E,t)$ were
calculated with weight coefficients $a$ in order that they can be compared
with theoretical ones. For $i$-th event in the sample of $n$ events
\begin{equation}
\label{eq3}
a_i=E^2_i f_i^{-1}(E_{+})(n^{-1}\sum_{i=1}^n E_i^{-2} f_i^{-1}(E_+))^{-1}
\end{equation}

Such structure of $a_i$ corresponds to the
process of inverse $\beta$-decay,
\begin{equation}
\label{eq4} \bar \nu_e + p \to n + e^+
\end{equation}

\noindent which, as it is assumed, causes the registration of
$\bar \nu_e$ in Cherenkov detectors. Really, process (\ref{eq4})
is 15-20 times more probable than the competing process of the
scattering of neutrinos (of all types) on electrons. Besides,
precisely (\ref{eq4}) allows one to compare the energy of neutrino
and the number of events observed in \cite{1,2} with predictions
of the stellar core collapse theoretical models. Formula
(\ref{eq3}) takes into account the quadratic growth of the capture
cross section (\ref{eq4}) with energy $E$ and registration
probability $f(E_+)$ of a positron with energy $E_+$ in a detector
(data of KII and IMB collaborations presented in \cite{5} were
used in computations).

Usually the relation between $E$  and $E_+$  is
expressed by formula

\begin{equation}
\label{eq5}
E=E_+ + 1.3\,MeV
\end{equation}

 In this work a more accurate relation~\cite{6},
which takes into account a small correction related to the
recoil energy of neutron is used. However, in the case when $E_+ > E_c$,
expression (\ref{eq5})
is sufficiently accurate and it is not necessary to use information
about the angular distribution of positron tracks in detectors for
process (\ref{eq4}).
If one considers elastic scattering of neutrinos on electrons it is
necessary to use such (very inaccurate) information.

The properties of the events from
neutrino burst in KII~\cite{1} and IMB~\cite{2} detectors are summarized in
the Tab.~\ref{tab1}.

\begin{table}[ht]
\begin{center}
\caption{Measured properties of the positron
events detected KII and IMB in the neutrino burst\textbf{$^{a}$}}
\small
\begin{tabular}{p{1.0cm}p{1.0cm}p{2.0cm}p{0.01cm}p{1.0cm}p{1.0cm}p{2.0cm}}
\multicolumn{7}{c}{} \\
\hline
\multicolumn{3}{c}{KII detector~\cite{1}} & &
\multicolumn{3}{c}{IMB detector~\cite{2}} \\
\cline{1-3}\cline{5-7}
Event
number\textbf{$^{b}$} &
Event
time $t_d$
(sec) &
Positron
energy $E_+$
(MeV) & &
Event
number &
Event time
$t_d$ (sec) &
Positron
energy $E_+$
(MeV)\\
\cline{1-3}\cline{5-7}
1  & 0      & 20.0 $\pm$ 2.9 & & 1 &   0  & 38 $\pm$ 9.5 \\
2  & 0.107  & 13.5 $\pm$ 3.2 & & 2 & 0.42 & 37 $\pm$ 9.3 \\
3  & 0.303  &  7.5 $\pm$ 2.0 & & 3 & 0.65 & 40 $\pm$ 10  \\
4  & 0.324  &  9.2 $\pm$ 2.7 & & 4 & 1.15 & 35 $\pm$ 8.8 \\
5  & 0.507  & 12.8 $\pm$ 2.9 & & 5 & 1.57 & 29 $\pm$ 7.3 \\
7  & 1.541  & 35.4 $\pm$ 8.0 & & 6 & 2.69 & 37 $\pm$ 9.3 \\
8  & 1.728  & 21.0 $\pm$ 4.2 & & 7 & 5.01 & 20 $\pm$ 5   \\
9  & 1.915  & 19.8 $\pm$ 3.2 & & 8 & 5.59 & 24 $\pm$ 6   \\
10 & 9.219  &  8.6 $\pm$ 2.7 & &   &      &            \\
11 &10.433  & 13.0 $\pm$ 2.6 & &   &      &            \\
12 &12.439  &  8.9 $\pm$ 1.9 & &   &      &            \\
\hline
\multicolumn{7}{l}
{
$^a$It is suggested that all events were caused by
the capture of $\bar \nu_e$.
}\\
\multicolumn{7}{l}
{
$^b$Event no.6 has been excluded by authors from
the signal analysis.
}
\end{tabular}
\end{center}
\label{tab1}

\end{table}

\begin{figure}[t]
\caption{%
The results of the statistical analysis of neutrino
events in IMB detector. (a) The dependence of variance $D(t)$ on $E_0$
(assumed rest energy of $\bar \nu_e$).
(b) The dependence
of the coefficients of correlation on $E_0$: $Q(E^{-2},t)$ -
curve 1 and $Q(E,t)$ - curve 2. Dashed lines limit FWHM of distributions
according to $Q(E,t)$ obtained by Monte Carlo method. The straight dotted
lines designate the strip of the model theoretical values of
$\langle Q(E,t)\rangle$ averaged with
respect to the neutrino burst.
}
\begin{center}
\includegraphics[width=4in,bb=0 0 600 600]{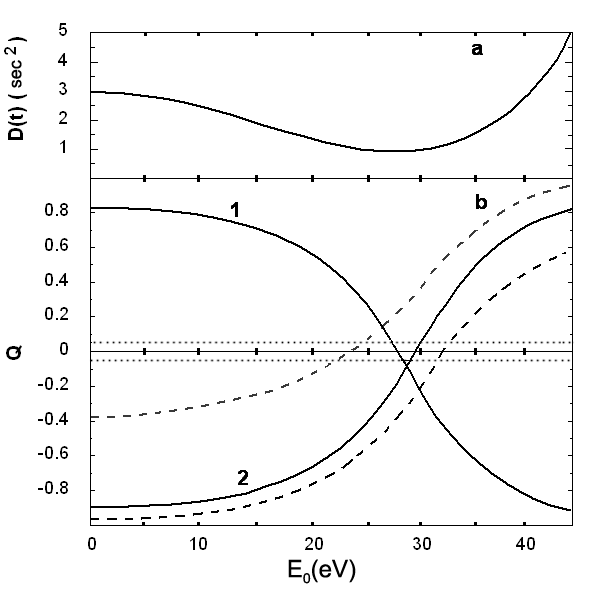}\\
\end{center}
\label{fig:1}
\end{figure}

\begin{figure}[t]
\caption{%
The same as in Fig.~\ref{fig:1} for neutrino events in KII
detector.
}
\begin{center}
\includegraphics[width=4in,bb=0 0 600 600]{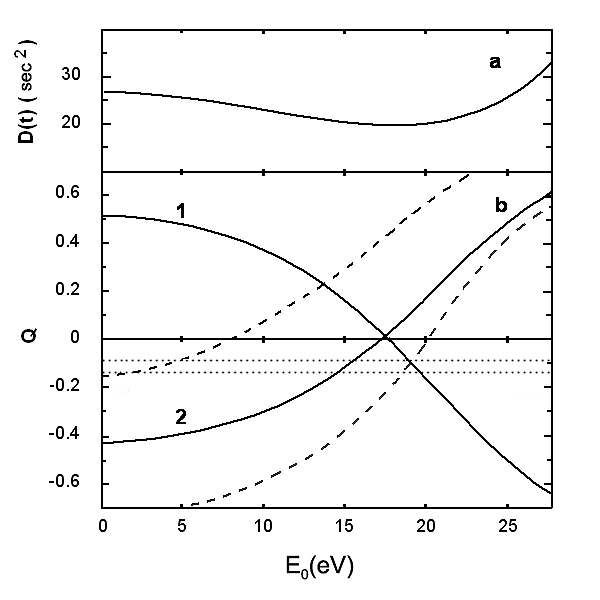}\\
\end{center}
\label{fig:2}
\end{figure}

The distance to SN1987A is probably $55\pm15\, Kpc$~\cite{6}. The value
$d=56\, Kpc$
was chosen as well as in~\cite{4}. When we carry out the statistical
analysis of neutrino events, it is significant that they all should be
caused by the capture of $\bar \nu_e$ (\ref{eq4}).
According to the estimates of authors of papers~\cite{1,2}, the
background count rate during 10~s was $\sim0.2$ (KII) and
$\sim0.8$ (IMB). One event in
IMB was eliminated by authors as a background one. I considered all
other 8 events. Amongst those eleven events presented by KII event no.
3 has value $E_+ = 7.5 \pm 2.0\,MeV$,
which practically coincide with energy threshold of KII
detector. With large probability this event may be considered as a
background one~\cite{6}, and it was not taken into account in our
computations. Thus, two statistical samples of events $(E,t_d)$
were taken as
initial data, one with size $n=10$ (KII) and another with
$n=8$ (IMB). Each sample
has its own function $F(E,t;E_c)$ as population, since values of
$E_c$  (as parameters)
for considered detectors differ essentially. According to (\ref{eq1}),
for each of the assumed values of $E_0$, the values $t_i$
were determined from $t_{di}$
 values recorded for considered detector, which gave the sample of $n$
events $(E_i, t_i)$ specifying density function
$F(E,t;E_c)$ of the source. Sample variance $D(t)$
and correlation coefficients $Q(E,t)$  and $Q(E^{-2},t)$ were computed.
Fig.~\ref{fig:1} and Fig.~\ref{fig:2}
represent the dependencies of these quantities on $E_0$.

Let us note that these sample moments are invariant
with respect to time shift. It is essential for our problem and makes
possible comparison of the results obtained from detectors which are
not synchronized in time. Let us emphasize also that considered
functions do not depend on astrophysical model (except the assumption
that process of capture (\ref{eq4})
is real). It is evident from figures that curves $D(t)$ have minima.

Is it possible to associate the location of the minimum
on $E_0$  scale with sought value of the $\bar \nu_e$ mass?
Let us consider the relation
between $D(t)$ on the one hand and $Q(E^{-2},t)$ and $Q(E,t)$
 on the other hand. It can be
demonstrated from (\ref{eq2}) that

\begin{equation}
\label{eq6}
D(t)=D(t_d)-\frac{1}{4} d^2c^{-2} \cdot D(E^{-2}) E_0^4-dc^{-1}\cdot Q(E^{-2},t)\cdot S(E^{-2})\cdot S(t)\cdot E_0^2
\end{equation}

\noindent
where $S(t)$ and $S(E^{-2})$ are the standard deviations of quantities $t$
and $E^{-2}$. The first term in (\ref{eq6})
is the variance of the observed values of $t_d$ (it coincides with $D(t)$ when
$E_0=0$),
the second term is proportional to the variance of $E^{-2}$ values, and
$Q(E^{-2},t)$
is the coefficient of correlation of $t$  and $E^{-2}$ variables
in the third term. If one uses sample values of statistical moments in
(\ref{eq6}), relation (\ref{eq6})
produces just the same values of $D(t)$ as a direct method presented above.
Expression (\ref{eq6})
allows to understand the behaviour of curve $D(t)$ in dependence on $E_0$.
When
values of $E_0$ are small, two last terms are negative $(Q(E^{-2},t)>0)$.
When $E_0$ increases,
the last term becomes positive, since inequality $(Q(E^{-2},t)<0)$
becomes correct.
Computations demonstrate that the minimum in considered curve precisely
corresponds to such value of $E_0$  where $Q(E^{-2},t)$ changes its sign.
In this region
$E_0$ both functions $Q(E^{-2},t)$ and $Q(E,t)$ have large $E_0$ derivative.
The aforesaid is illustrated by Fig.~\ref{fig:1} and Fig.~\ref{fig:2}. It is also evident from
figures that standard correlation coefficient $Q(E,t)$ changes its sign at
approximately the same value of $E_0$ as $Q(E^{-2},t)$. Just function $Q(E,t)$
is considered
below. It is evident from Fig.~\ref{fig:1} and Fig.~\ref{fig:2} that functions $Q(E,t_d)$
corresponding
to direct observations either are close to unity with respect to
modulus (IMB) or have the same order of magnitude (KII), i.e. there
evident correlation "$E-t_d$" exist. Function $Q(E,t_d)$ specifies "E-time" correlation 
near detector. Only
modern astrophysical models of stellar core collapse can present
information concerning $Q(E,t)$ near the source.

The results of~\cite{4} were used in order to compute
coefficient of correlation $\langle Q(E,t)\rangle$ averaged with respect to neutrino
burst. In~\cite{4} the equations of hydrodynamics (in the framework of the general
theory of relativity) and equations of multigroup (with respect to
energy) transport of neutrinos which takes into account all types of
neutrinos and antineutrinos were solved jointly numerically. In
particular in~\cite{4} luminosity for $\bar \nu_e$ and mean instantaneous
(at $t=const$)
energy of $\bar \nu_e$ are represented as functions of time which are necessary for
computation of $\langle Q(E,t)\rangle$.
At the stage of the core cooling $(t \ge 1.2 s)$ exponential decrease
of the neutrinosphere temperature was assumed with time constant
$\tau_{\scriptscriptstyle T}=5.6 s$
in accordance with graphical data~\cite{4}. Momentary energy spectra of
$\bar \nu_e$  were
approximated by Fermi-Dirac (F-D) distributions. Such spectra which
were computed in~\cite{4} differ from F-D spectra. They demonstrate
steeper decrease at high energies of $\bar \nu_e$.
The model energy spectra of $\bar \nu_e$
with the same peculiarity at high energies~\cite{7} were used in one
version of our computations. In this case resulting values of $\langle Q(E,t)\rangle$
differ
less than to 10\%. Possibly it is caused by the fact that mean
momentary energies of $\bar \nu_e$ were fixed in computations according
to data of~\cite{4}. The following approach was used in order to compute the model
theoretical value of $\langle Q(E,t)\rangle$  in the same manner as in
the case of the sample
empirical values of $Q(E,t)$. Values of $E_i$ and $t_i$
were computed by Monte Carlo
method in order to obtain samples of size $n$ (as a rule, these sizes
coincided with empirical ones). Further value of
$\langle Q(E,t)\rangle$ was computed for each $k$-th sample, and averaging
with respect to $k (k_{max}=500)$ was carried out. Obtained
strips of values of $\langle Q(E,t)\rangle$ limited by point straight
lines as a result of the
variation of energy spectra of $\bar \nu_e$ and sizes $n$
are presented in Fig.~\ref{fig:1} and Fig.~\ref{fig:2}.
It is evident from figures that modulus of $\langle Q(E,t)\rangle$
is much less that unity,
i.e. the level of "$E-t$" correlations in neutrino burst is small on the
average. This is caused by a large role of accretion phase in the
process of gravitational collapse of stellar core. For cores with
masses $\approx 2M_{\odot}$ more than half of $\bar \nu_e$ are
emitted at the stage of accretion specified by positive sign of $Q(E,t)$~\cite{4,5}.
Phase of cooling corresponds to negative values of $Q(E,t)$. Thus both
phases to some extent compensate each other in computation of $\langle Q(E,t)\rangle$.
High
energy threshold $E_c$ of neutrino detection increases the role of accretion
phase since $\bar \nu_e$ with higher energies are emitted at this phase.
Values $E_c=20\,MeV$
(IMB) and $E_c=8\,MeV$ (KII) were used in computations.
Thus one can understand a
positive gap between the values of $\langle Q(E,t)\rangle$ for
IMB on the one hand and of KII
on the other hand (see Fig.~\ref{fig:1} and Fig.~\ref{fig:2}). Large (with respect to modulus)
and negative (with respect to the sign) differences
$(Q(E,t) - \langle Q(E,t)\rangle)$ specifying
neutrino events in two detectors, evidence that observed coefficients
of correlation $Q(E,t_d)$ are mainly related to the effect of the delay of the
neutrinos with less energy according to (\ref{eq1}),
i.e. suggest nonzero mass of $\bar \nu_e$.

Determining the curves $Q(E,t)$ intersection regions (as
functions of $E_0$) with strips $\langle Q(E,t)\rangle$
we arrive at the estimates of $E_0$ as $29 \div 31\,eV$ for IMB
and $14 \div 16\,eV$ for KII. If we assume that the accuracy of IMB and KII
measurements are equal, $E_0$ of electron antineutrino can be estimated as

\begin{equation}
\label{eq7}
E_0=22 \pm 10\,eV
\end{equation}

Here presented error is the root-mean-square deviation
for the sample of the size $n=2$. Assuming that this error has random
nature, let us employ Student's distribution with degrees of freedom
$m=1$
for the estimation of the probability of the assumption that $E_0 < 2 \, eV$
is valid.
This probability does not exceed $0.15$. Thus we can discard this
hypothesis at 85\% significance level.

If we want to answer the question formulated above, one
may say that the minimum of $D(t)$ (according to $E_0$ scale)
corresponds to the
region of the possible values of $E_0$ since this minimum is related to the
low level of "$E-t$" correlations.

\begin{figure}
\caption{%
Distributions of $Q(E,t)$
computed by Monte Carlo method for detectors IMB (a) and KII (b). The
errors of determination of the energy of positrons in detectors are
taken into account. Thin lines represent distribution for
$E_{0}=0 \div 2\,eV$, thick lines
- for $E_0=22\,eV$. The regions of the
overlapping of histograms are dashed.
}
\begin{center}
\includegraphics[width=3in,bb=0 0 575 750]{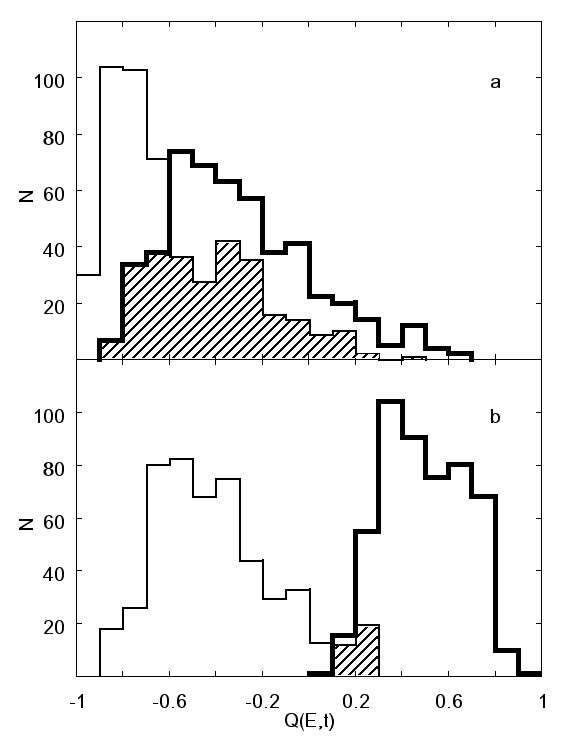}\\
\end{center}
\label{fig:3}
\end{figure}

Aside from random errors,
"true" value of $E_0$ can be distorted
by initial assumption that all neutrino events are caused by process of
capture (\ref{eq4}). In~\cite{6} an assumption was considered that event no. 4
in KII was
caused by the scattering of neutrino on electron. If one takes account
of such possibility, it complicates the analysis since in particular it
requires the construction of function $F(E,t;E_c)$ including not only
$\bar \nu_e$, but all
other types of neutrinos and antineutrinos. Nevertheless let us
consider for illustration how the values of $D(t)$ and $Q(E,t)$ vary
(depending on $E_0$)
if we interpret event no.4 (KII) as the scattering of $\bar \nu_e$.
Considering
gravitational collapses of stellar cores one limits the energy of
emitted neutrinos by $\sim 50\,MeV$. In this case the recoil
electrons with energy exceeding $E_c$ should be scattered through angles
$\varphi \leq 20^{\circ}$. If to assume that
$\varphi = 10^{\circ}$ for event no.4, we arrive at $E=29\,MeV$ (instead
of $10.6\,MeV$ in the case of capture). Such value of
$E_i (i=4)$ leads to essential variation of  the $D(t)$ and $Q(E,t)$
values. In particular, $D(t_d)=24.6(27)$,
minimum value of the variance $D(t_{min})=9(20)$
and  $Q(E,t_d) = -0.68 (-0.42)$. The numbers in
parentheses correspond to the process of the capture. The intersection
of $E_{0}$ axis with curve $Q(E,t)$ takes place at the value of $E_0$
which is larger
by $3\,eV$ than the value represented in Fig.~\ref{fig:2}. For the estimation of
$E_0$ in this case we obtain

\begin{equation}
\label{eq8}
E_0=24 \pm 8\, eV
\end{equation}

Considered example demonstrates that taking into
account the possibility $\bar \nu_e$ scattering, allows one to make more similar
the results obtained by data from both detectors, however it does not
change essentially the estimate of $\bar \nu_e$ mass.

We used errors $\delta E$ (standard deviations)~\cite{1,2} in
order to take into account the influence of inaccuracies in the
determination of positrons $E_+$  energy in Cherenkov detectors. We computed
the values $\tilde E_{+i}$ for Gaussian distribution with prescribed value
of $\delta E_{+i}$ and $M \tilde E_{+i} = E_{+i}$ by
Monte Carlo method. The results of computations were the samples with
sizes $n$ ( ($n=10$) (KII) and  $n=8$ (IMB))
for $\tilde E_+$ and $\tilde E$. The coefficient of correlation
$Q(\tilde E,t)$
for a number of values of $E_0$ was computed in each sample of values of $\tilde E$.
A number of $k$ samples $(k=500)$ were used for a fixed value of $E_0$
in order to obtain
a distribution (histogram) according to the value of $Q$. Examples of such
distributions are presented in Fig.~\ref{fig:3}.

Dashed lines in Fig.~\ref{fig:1} and Fig.~\ref{fig:2} limit FWHM of these
distributions (smoothed histograms). It is evident that the
distributions have sufficiently large FWHM. The point is: how
significant is the difference of the values of $Q(E,t)$ for
$E_0=0 \div 2\,eV$ (the carried out
analysis cannot distinguish $E_0$ in this interval) and for the value of
$E_0=22\,eV$
estimated the mass of $\bar \nu_e$ by the method presented above?
Fig.~\ref{fig:3} represents
distributions with respect to $Q$  for $E_0=22\,eV$ and similar
distributions for $E_0=0 \div 2\,eV$. The
regions of overlapping allow one to estimate the probability $W$ of
coincidence of the distributions for $E_0=0 \div 2\,eV$ and $E_0=22\,eV$
in the following manner: $W(IMB) \cong 0.54$
and $W(KII) \cong 0.07$. The probability of the fact that despite the data of
both detectors (independent measurements), distributions for
$E_0=0 \div 2\,eV$ and $E_0=22\,eV$ do not
differ is

$$W=W(IMB)\times W(KII) \cong 0.04.$$

Thus the statement that our analysis cannot distinguish
$E_0=0 \div 2\,eV$ and the estimate of $E_0$ in (\ref{eq7})
can be rejected at significance level not less than 90\%.

\begin{figure}
\caption{%
Integral rate of neutrinos for detector IMB: smooth
curve corresponds to theoretical model~\cite{4}, step spectra correspond
to real neutrino events (dotted line corresponds to $E_{0}=0\,eV$,
solid line -- to $E_{0} = 30\,eV$).
}
\begin{center}
\includegraphics[width=3in,bb=0 0 450 380]{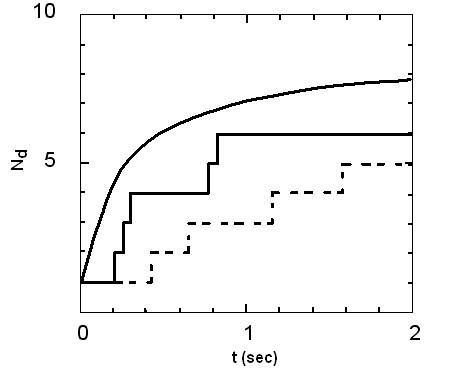}\\
\end{center}
\label{fig:4}
\end{figure}

In the analysis presented above we used data~\cite{4}
related to an iron stellar core with mass
$2M_{\odot}$. Smooth curve in Fig.~\ref{fig:4}
represents integral neutrino event rate for IMB detector which was theoretically
predicted~\cite{4}.  Step integral spectrum represented by dotted line
corresponds to experimental data. It is evident that these spectra do
not agree well. On this basis author of~\cite{4} concluded that one
should pass to the core of different mass ($1.35 M_{\odot}$). Indeed,
with the employment of Kolmogorov-Smirnov criterion one can reject
theoretical model for $2M_{\odot}$ at significance level
$\sim 99\%$. However, the time delay of slow neutrinos with
respect to the fast ones on the path from the source to the detector is
not taken into account in~\cite{4}, i.e. the case
$E_0=0$ was implicitly considered. Solid
step lines in the same figure represent integral spectrum for
$E_0=30\,eV$. This spectrum much
better agrees with theory and one can demonstrate that in this case
theoretical model of the core with mass
$2M_{\odot}$ cannot be rejected with usually used
significance levels. Let us note that value
$E_0=30\,eV$ determined by
fitting from the condition of best agreement with theoretical curve
coincide with value obtained by the method presented above.

It is interesting to compare estimates (\ref{eq7})
and (\ref{eq8})
with experimental results where the mass of $\bar \nu_e$
was determined from data
of tritium $\beta$-decay. New result of these experiments is the discovery of
anomalous structure of $\beta$-spectra~\cite{8,9} -- a very weak line
near end point of $\beta$-spectrum. In~\cite{10,12} an approach was proposed which
takes into account the anomalous structure and leads to the value $E_0=20 \pm 5\,eV$.

This result agrees with estimates (\ref{eq7})
and (\ref{eq8}).

\section{Conclusions}

Modern model of the gravitational stellar core collapse
with an accretion phase predicts the low level of correlation between
the energies $E$ of electron antineutrinos $\bar \nu_e$
emitted by the star and the
ejection instants $t$ of $\bar \nu_e$ on the average.

Similar coefficients $Q(E,t)$ of
"$E-t$" correlation can be calculated for neutrino events
recorded by Cherenkov water detectors of KII and IMB collaborations
from SN1987A neutrino burst.

These values $Q(E,t)$ depend on the assumed mass of $\bar \nu_e$.
On condition that empirical values $Q(E,t)$ must be equal to the
theoretical model quantities $\langle Q(E,t)\rangle$ averaged with
respect to neutrino burst it can obtained the value
$\sim20\,eV/c^{2}$ as an estimate of the nonzero $\bar \nu_e$
mass.

The error of this estimate allows to consider that the $\bar \nu_e$
mass less than $2\,eV/c^{2}$ can be realized only with little
likelihood.

There are two ways of the $\bar \nu_e$ mass estimation: one from
the astrophysical observations by the method presented above, and the
other from laboratory data of the tritium
$\beta$-decay. Both
methods agree adequately when taken into account the anomalous
structure near end point of the
$\beta$-spectrum.

\section*{Acknowledgements}{
The author is grateful to G. T. Zatsepin, L. I.
Sarycheva, B. A. Khrenov and S. I. Svertilov for a fruitful
discussion.
}


\begin{thebibliography}{99}

\bibitem{1} K. Hirata et al. Phys. Rev. Lett {\bf 58}, 1490 (1987)

\bibitem{2} R.M. Bionta et al. Phys. Rev. Lett {\bf 58}, 1494 (1987)

\bibitem{3} G.T. Zatsepin. Pis'ma Zh. Exp. Teor. Fiz {\bf 8}, 333 (1968)

\bibitem{4} S.W. Bruenn. Phys. Rev. Lett. {\bf 59}, 938 (1987)

\bibitem{5} D.N. Schramm and J.W. Truran. Physics Reports {\bf 189}, 89 (1990)

\bibitem{6} E.W. Kolb, A.J. Stebbins, M.S. Turner.  Phys. Rev. D {\bf 35}, 3598 (1987)

\bibitem{7} D.K. Nadezhin and I.V. Otroshchenko. Astron.Zh. {\bf 57}, 78 (1980)

\bibitem{8} W. Stoeffl, D.J. Decman. Phys. Rev. Lett. {\bf 75}, 3237 (1995)

\bibitem{9} A.I. Belesev et al. Phys. Lett. B {\bf 350}, 263 (1995)

\bibitem{10} B.I. Goryachev. Bulletin of the Lebedev
Physics Institute {\bf 3}, 26 (2003); Kratkie Soobsheniya po
Fizike {\bf 3}, 33 (2003)

\bibitem{12} B.I. Goryachev. Preprint of INPH of
MSU {\bf 41/681} (2001)

\end{thebibliography}
\end{document}